\documentclass{article}

\usepackage{arxiv}

\usepackage[utf8]{inputenc} 
\usepackage[T1]{fontenc}    
\usepackage{hyperref}       
\usepackage{url}            
\usepackage{booktabs}       
\usepackage{amsfonts}       
\usepackage{nicefrac}       
\usepackage{microtype}      
\usepackage{lipsum}
\usepackage{graphicx}
\graphicspath{ {./images/} }

\usepackage{subcaption}
\usepackage{bbm}
\usepackage{algorithm}
\usepackage{algpseudocode}
\usepackage{amsmath}
\usepackage[numbers]{natbib}
\usepackage[numbers]{natbib}
\newcommand{\ind}{\mathbbm{1}}

\title{Structured Gaussian Processes for Uncertainty-Aware Classification of High-Dimensional, Small-Sampled Omics Data}

\author{
 Yue Zhang \\
  Department of Mathematical Sciences\\
  Durham University\\
  Durham, United Kingdom \\
  \texttt{yue.zhang@durham.ac.uk} \\
   \And
 Nandini Amit Gadhia \\
  STFC Hartree Centre\\
  University of Liverpool\\
  Liverpool, United Kingdom \\
  \texttt{Nandini.gadhia@stfc.ac.uk} \\
  \And
 Georgios Karagiannis \\
  Department of Mathematical Sciences\\
  Durham University\\
  Durham, United Kingdom \\
  \texttt{georgios.karagiannis@durham.ac.uk} \\
  \And
 Michalis Smyrnakis \\
  STFC Hartree Centre\\
  Daresbury, United Kingdom \\
  \texttt{michail.smyrnakis@stfc.ac.uk}\\
}

\begin{document}
\maketitle
\begin{abstract}
Classifying heterogeneous omics data remains a fundamental challenge in computational biology, particularly in high-dimensional, small-sample settings where nonlinear interactions dominate and class imbalance further complicates reliable prediction of minority phenotypes. While traditional kernel methods rely on feature abundance, they fail to leverage the known interaction landscapes of biological systems.

In this work, we propose a structured Gaussian process classification framework that integrates graph-encoded biological pathways directly into the kernel construction. By propagating information along known interaction networks and combining this with abundance-derived features, the resulting classifier captures both quantitative measurements and topological context.

We benchmark our proposed methodology on three publicly available (gut and fecal) microbiome datasets. To address severe class imbalance, we evaluate complementary strategies, including data-level resampling, threshold calibration, and confusion-matrix–based adjustments, and report minority-class performance alongside accuracy. The hybrid approach yields a performance gain over unstructured baselines and matches the performance of established benchmarks for similar datasets. Furthermore, the probabilistic nature of the framework naturally provides calibrated predictive uncertainty, enabling robust differentiation between confident predictions and ambiguous samples.
\end{abstract}


\section{INTRODUCTION}

High-dimensional omics datasets such as microbiome profiles are increasingly used for disease classification and biomarker discovery. In practice, these tasks are challenging because cohort sizes are often small relative to the number of measured features and class imbalance can be severe. In these settings, models relying solely on point predictions are often unreliable, making it difficult to separate high-confidence results from ambiguous cases that require further review.

Probabilistic classification provides a natural way to address these issues \cite{hart2001pattern}. Gaussian process classification (GPC) is particularly attractive in small-sample settings because it is Bayesian and non-parametric, producing predictive probabilities together with uncertainty estimates \cite{williams2006gaussian}. However, the practical performance of GPC depends critically on the kernel \cite{williams2006gaussian}, which defines the similarity between samples. We found that standard kernels built purely from abundance vectors ignore known biological structure, even though interactions often provide informative context for prediction. In our experiments, a vector-only GP baseline can have limited discrimination on some cohorts under severe imbalance, which motivates incorporating structured information into the kernel.

In this work we develop a structured Gaussian process classification framework for uncertainty-aware prediction on small-sample, high-dimensional omics data. Our key idea is to incorporate graph-encoded biological knowledge directly into the kernel that defines sample similarity. For each sample, we represent biological relationships as a weighted interaction network and propagate node-level signals to obtain a structure-aware embedding \cite{neumann2016propagation}. We then combine a propagation-based graph kernel with a standard kernel on abundance-derived vector features, yielding a hybrid kernel that can adaptively balance quantitative measurements and topological context \cite{williams2006gaussian}.

We perform Bayesian inference using the Laplace approximation and, where appropriate, sampling over kernel hyperparameters to propagate uncertainty into predictions. Beyond overall accuracy, we focus on minority-class performance and investigate complementary strategies for handling imbalance, including resampling, threshold calibration, and a confusion-matrix-based prior adjustment.

We evaluate our proposed approach on three available microbiome datasets and compare against unstructured baselines. Our main contributions are a graph-informed hybrid kernel for GP classification, a systematic evaluation of imbalance strategies within this framework, and empirical results on three microbiome datasets including calibration and uncertainty analyses.

The remainder of the paper is organised as follows. Section~\ref{sec:background} reviews background, Section~\ref{sec:method} presents our proposed method, Section~\ref{sec:imbalance} describes the imbalance-handling strategies, Section~\ref{sec:results} presents numerical experiments and results, and Section~\ref{sec:conclusion} concludes.  

\section{BACKGROUND}
\label{sec:background}

The human microbiome is critical to host physiology and immune function, with dysbiosis in omics layers typically linked to phenotype presentation. However, establishing robust associations is complicated by the complexity of the ecosystem. This is particularly exacerbated in gut and fecal microbiome experiments, where exogenous factors such as diet and host genetics heavily influence metabolic profiles. The resulting data are high-dimensional, compositional, and inherently nonlinear, complicating traditional modelling approaches. Gaussian Processes (GPs) offer a powerful, non-parametric model for Bayesian inference in this setting.  By modelling nonlinear structure directly and offering principled uncertainty quantification, GPs enable robust disease stratification while maintaining interpretability in matched microbiome–metabolome datasets. 

\subsection{Gaussian process classification}

Gaussian processes (GPs) regression provides  a flexible Bayesian framework for supervised learning \cite{williams2006gaussian}. Although standard GPs are typically associated with regression tasks where output is continuous, Gaussian process classification (GPC) extends this framework to model categorical output.

Consider a binary classification task with a dataset of $N$ observations $\mathcal{D}=\{(\mathbf{x}^{(i)}, y^{(i)})\}_{i=1}^{N}$, where each input is a $d$-dimensional vector $\mathbf{x}^{(i)}\in\mathbb{R}^d$ and the corresponding class label is $y^{(i)}\in\{0,1\}$. 
For notational convenience, we denote the inputs in a design matrix $\mathbf{X} = [\mathbf{x}^{(1)}, \dots, \mathbf{x}^{(N)}]^\top \in \mathbb{R}^{N \times d}$ and the targets in a vector form $\mathbf{y} = [y^{(1)}, \dots, y^{(N)}]^\top$. Since the outputs are binary, we assume the class labels follow a Bernoulli distribution conditionally independent given a probability of success parameter $\pi^{(i)}$
\begin{equation}
    y^{(i)} \mid \pi^{(i)} \sim \mathrm{Bernoulli}\big(\pi^{(i)}\big), \quad i = 1, \cdots, N.
\end{equation}
Consider a latent function $f: \mathbb{R}^d \to \mathbb{R}$ that models the underlying relationship in the input space. The probability of success $\pi^{(i)}$ is related to this unobserved latent function value $f(\mathbf{x}^{(i)})$ through a sigmoid function $\sigma(\cdot)$ that maps the real-valued output into the unit interval $[0,1]$
\begin{equation}
    \pi^{(i)} = \sigma\big(f(\mathbf{x}^{(i)})\big).
\end{equation}
Common choices for $\sigma(\cdot)$ include, for example, the logistic function $\sigma(z) = (1+e^{-z})^{-1}$ and the standard Gaussian cumulative distribution function (CDF) $\sigma(z)=\Phi(z)$.
    
To capture the uncertainty about the unknown function $f$, a Gaussian process prior can be assigned over the latent function as 
\begin{equation}
    f(\cdot) \sim \mathcal{GP}\big(\mu(\cdot),  k(\cdot,\cdot)\big), 
\end{equation}
where $k(\cdot,\cdot)$ is a positive semi-definite covariance function  that encodes assumptions about the smoothness and structure of the function and $\mu(\cdot)$ is the mean function often set as zero \cite{williams2006gaussian}.

The choice of the covariance function (or kernel) is central to GP models, as it encodes assumptions about the smoothness and structure of the latent function $f(\cdot)$. A kernel $k: \mathcal{X} \times \mathcal{X} \to \mathbb{R}$ is typically stationary if it depends only on the distance. A flexible and widely used stationary kernel is the Mat\'{e}rn covariance function \cite{matern1986spatial}:
\begin{equation}
    k_{\text{Mat\'ern}}(d; \boldsymbol{\theta}) = \sigma^2\frac{2^{1-\nu}}{\Gamma(\nu)} \left( \sqrt{2\nu} \frac{d}{l} \right)^\nu K_\nu \left( \sqrt{2\nu} \frac{d}{l} \right),
\end{equation}
where $d$ is the distance, $\boldsymbol{\theta} = \{l,\sigma\} $ are the hyperparameters, $K_\nu$ is the modified Bessel function of the second kind, and $\nu > 0$ is a parameter that controls the smoothness of the resulting GP; larger values of $\nu$ correspond to smoother sample paths. As $\nu \to \infty$, the Mat\'{e}rn covariance function reduces to the widely used Gaussian kernel (also known as the radial basis function (RBF) kernel)
\begin{equation}
    k_{\text{SE}}(\mathbf{x}, \mathbf{x}') = \sigma^2 \exp\left(-\frac{\|\mathbf{x} - \mathbf{x}'\|^2}{2l^2}\right),
\end{equation} 

For Bayesian inference over the kernel hyperparameters $\boldsymbol{\theta}$, we place prior distributions on these parameters. Common choices for positive-valued hyperparameters such as the length-scale $l$ and signal variance $\sigma^2$ include the inverse-gamma distribution $\mathcal{IG}(\alpha,\beta)$, and the gamma distribution $\mathcal{G}(\alpha,\beta)$.

Predictive inference of the outcome $y_{*}$ at any point $\mathbf{x}_*$ is obtained by computing the predictive probability 
\begin{equation}
    p(y_* = 1 \mid \mathbf{x}_*, \mathbf{X}, \mathbf{y})
    = \int \sigma(f_*) \, p(f_* \mid \mathbf{x}_*, \mathbf{X}, \mathbf{y}) \, df_* ,
\end{equation}
that requires integration of $\sigma(f_*)$ with respect to the posterior of the latent function $f_*:=f(x_{*})$ at $x_{*}$. As the required integral is intractable, approximate computational methods are required, such as the Laplace approximation, expectation propagation (EP) or Markov chain Monte Carlo (MCMC) \cite{nickisch2008approximations}.  

Standard Gaussian process classification cannot be directly implemented in our problem, in which the input involves a graphical structure, and hence we properly extend this framework in Section \ref{sec:ModelSpecification}.


\subsection{Propagation kernel}
\label{sec:rw_kernel}

A key ingredient of our approach is a kernel that can compare graphs in a computationally simple way. We use information propagation on the graph to build such a similarity measure.

A graph $G=(V,E)$ consists of a set of nodes $V$ and edges $E\subseteq V\times V$. Its structure is represented by an adjacency matrix $\mathbf{A}\in\mathbb{R}^{m\times m}$ (with $m=|V|$), where $A_{ij}>0$ indicates an edge between nodes $i$ and $j$. The degree matrix is $\mathbf{D}=\mathrm{diag}(\mathbf{A}\mathbf{1})$, and the (unnormalised) graph Laplacian is $\mathbf{L}=\mathbf{D}-\mathbf{A}$, which characterises local smoothness of functions on the graph \cite{lauritzen1996graphical}. To capture global structural relationships, we use diffusion through a random walk on the graph \cite{hart2001pattern}. The transition matrix is
\begin{equation}
 \mathbf{T} = \mathbf{D}^{-1} \mathbf{A},
\end{equation}
where $T_{ij}$ is the probability of moving from node $i$ to node $j$ in one step. Let $\mathbf{f}^{(0)} \in \mathbb{R}^m$ be the vector of initial node features (e.g., node attributes or a uniform distribution vector). The propagation process updates the feature vector at each iteration $t$:
\begin{equation}
 \mathbf{f}^{(t+1)} = \mathbf{T} \, \mathbf{f}^{(t)},
 \qquad t = 0,1,\dots,t_{\max}-1.
 \label{eq:propagation_review}
\end{equation}
The vector $\mathbf{f}^{(t)}$ captures the accumulated information after $t$ steps of random walk propagation, integrating both initial node attributes and multi-hop structural relationships. The intermediate distributions $\{\mathbf{f}^{(t)}\}_{t=0}^{t_{\max}}$ already encode structural information at multiple scales. Graphs with similar topology tend to induce similar node distributions at each propagation step. Propagation kernels \cite{neumann2016propagation} build graph similarity by comparing the propagated node-level distributions at each step and aggregating these similarities across $t$. In their general form, this involves defining a node kernel and summing comparisons over node pairs, typically together with binning or hashing of node representations to make computation feasible. 

However, this full propagation-kernel construction  requires additional design choices (e.g., hashing/binning schemes for node distributions) and introduces extra hyperparameters that are hard to tune in small-sample microbiome cohorts. For these reasons, we adopt a simplified propagation-based similarity that preserves the key idea (diffusion captures structure) while remaining compatible with standard GP kernels. Specifically, we use the propagated signal after $t_{\max}$ steps as a fixed-dimensional graph embedding, and measure similarity between two graphs using an RBF kernel on the embedding distance. This choice retains the diffusion intuition from propagation kernels while providing a straightforward kernel suitable for GP classification. The resulting graph-informed kernel is combined with a standard vector kernel in Section~\ref{sec:ModelSpecification}.

\section{METHODOLOGY}
\label{sec:method}


\subsection{GP classification with a hybrid kernel}
\label{sec:ModelSpecification}

We consider a supervised classification setting where each sample consists of a numerical feature vector and an associated graph. Formally, the training dataset is given by
\begin{equation}
\mathcal{D} = \{ (\mathbf{x}_i, G_i, y_i) \}_{i=1}^N, \quad \text{for } i=1,\dots,N,
\end{equation}
where $\mathbf{x}_i \in \mathbb{R}^d$ denotes a $d$-dimensional feature vector, $G_i = (V_i, E_i)$ is a sample-specific graph, and $y_i \in \{0,1\}$ is the corresponding class label. The graphs $G_i$ are undirected and weighted, with an adjacency matrix $A_i \in \mathbb{R}^{|V_i|\times |V_i|}$ encoding pairwise relationships between nodes (e.g., microbial species). Node-level attributes are represented by initial feature vectors $\mathbf{f}^{(0)} \in \mathbb{R}^{|V_i|}$, which correspond to the features associated with the nodes of graph $G_i$.

The binary outputs $\mathbf{y}$ are linked to the latent function values $f_i = f(\mathbf{x}_i, G_i)$ through a Bernoulli sampling distribution :
\begin{equation}
p(y_i \mid f_i) = \sigma(f_i)^{y_i} (1-\sigma(f_i))^{1-y_i}
\end{equation}
for $i=1,\dots,N$, where $\sigma(\cdot)$ is the logistic function $\sigma(z) = (1+e^{-z})^{-1}$. We assign a Gaussian process prior on a latent function $f(\cdot)$ mapping each sample $(\mathbf{x},G)$ to a real value:
\begin{equation}
f(\cdot) \sim \mathcal{GP}(0, k(\cdot,\cdot)),
\end{equation}
where $k(\cdot,\cdot)$ is a kernel that measures similarity between pairs of samples, integrating both numerical and graph-based information.

The key modelling task is therefore to construct a kernel function $k(\cdot,\cdot)$ that effectively incorporates the heterogeneous information present in the data. A kernel built only from $\mathbf{x}$ ignores biological relationships encoded in $G$, whereas a kernel built only from graphs discards potentially informative abundance-derived features. This motivates a hybrid kernel that combines both sources of information:
\begin{equation}
k_{\text{final}}((\mathbf{x}_1,G_1), (\mathbf{x}_2,G_2)) 
= k_{\text{vec}}(\mathbf{x}_1,\mathbf{x}_2) 
+ \lambda \, k_{\text{graph}}(G_1,G_2),
\label{GPC_Hkernel}
\end{equation}
where $k_{\text{vec}}$ is a standard kernel on vector features, $k_{\text{graph}}$ is a kernel measuring graph similarity, and $\lambda\ge 0$ controls the contribution of the graph component. The weight $\lambda$ is treated as an unknown parameter and is estimated from the data alongside other kernel hyperparameters. In the next subsection, we describe how $k_{\text{graph}}$ is constructed via propagation.

\subsection{Diffusion-based graph covariance function}

Our goal is to define a kernel that compares two sample-specific graphs in a way that captures both topology and node attributes, while remaining simple to compute. Following the diffusion intuition in propagation kernels \citep{neumann2016propagation}, we obtain a fixed-dimensional representation of
each graph by propagating an initial node signal through a random walk.
Moreover, we only consider the final propagated vector as input to an RBF kernel, unlike the original framework where histogram-based aggregations are aggregated across all iterations; this yields a simpler and more computationally efficient formulation. 


Consider an undirected weighted graph 
\begin{equation}
G = (V,E), \quad |V| = m,
\end{equation}
with an adjacency matrix $\mathbf{A} \in \mathbb{R}^{m \times m}$. Each node $u \in V$ is associated with a scalar attribute $f_u^{(0)} \in \mathbb{R}$, 
and we define $\mathbf{f}^{(0)} = [f_1^{(0)}, \dots, f_m^{(0)}]^\top \in \mathbb{R}^m$ 
as the vector of all node attributes. This initial signal typically corresponds to node-level features (e.g., normalised abundances in our application).

To encode the graph structure, we propagate node attributes across 
neighbourhoods via the transition matrix $\mathbf{P} = \mathbf{D}^{-1}\mathbf{A}$, 
where $D_{uu} = \sum_{v=1}^m A_{uv}$. Iterating for $t_{\max}$ steps,
\begin{equation}
    \mathbf{f}^{(t+1)} = \mathbf{P}\,\mathbf{f}^{(t)}, 
    \qquad t = 0,1,\dots,t_{\max}-1,
    \label{eq:propagation}
\end{equation}
yields the graph-level embedding
\begin{equation}
    \mathbf{g} = \mathbf{P}^{t_{\max}} \mathbf{f}^{(0)} \in \mathbb{R}^m,
    \label{eq:graph_embedding}
\end{equation}
which captures both initial node attributes and multi-hop structural information.

Directly computing similarity between complex, variable-sized graph structures is computationally challenging. By transforming each graph $G$ into a continuous, fixed-dimensional vector $\mathbf{g}$ in Euclidean space, we simplify the problem: the similarity between two graphs $G_i$ and $G_j$ can be efficiently measured by the distance between their corresponding embeddings $\mathbf{g}_i$ and $\mathbf{g}_j$.

Given two graphs $G_i$ and $G_j$, with propagated embeddings $\mathbf{g}_i$ and $\mathbf{g}_j$ as in~\eqref{eq:graph_embedding}, their similarity can be defined using a Gaussian radial basis function (RBF) kernel:
\begin{equation}
    k_{\text{graph}}(G_i, G_j) 
    = \sigma_g^2\exp\!\left( - \frac{\|\mathbf{g}_i - \mathbf{g}_j\|_2^2}{2\ell_g^2} \right),
    \label{eq:graph_kernel}
\end{equation}
where $\sigma_g^2>0$ controls the marginal variance contributed by the graph component and $\ell_g>0$ is a length-scale parameter. This kernel measures graph similarity through the Euclidean distance between their diffusion-based embeddings, assigning larger covariance to graphs with
more similar propagated representations. This graph covariance function is then combined with a standard vector kernel to form the hybrid kernel in \eqref{GPC_Hkernel}. We next describe how the sample-specific graphs $G_i$ are constructed from the omics data, which provides the input for the diffusion step used to compute the embeddings $\mathbf{g}_i$.

\subsection{Graph construction}

The following section details two methods for sample-specific graph construction, optimized for single-cell omics data.

\begin{itemize}
    \item \textbf{LIONESS (Pearson Correlation Networks)}: LIONESS (Linear Interpolation to Obtain Network Estimates for Single Samples) is a framework for estimating sample-specific networks by decomposing an aggregate, population-level network \citep{kuijjer2019lioness}. It utilises Pearson correlation to infer dense, weighted symmetric networks. 
    
\item \textbf{Weighted co-occurence consensus networks}:
To capture both direct and indirect interactions, as well as shared responses to abiotic effects, we can also construct sample-specific networks by using a ratio-based edge definition coupled with bootstrap filtering \cite{gadhia2024novelapproachdifferentialexpression}. The edge weighting in this methodology emphasizes species that change congruently, regardless of their absolute abundance magnitude. Weights approaching 1 indicate abundance mutualism, whereas lower weights suggest commensalism or parasitism.
\end{itemize}

With the graphs constructed, we can compute $k_{\text{graph}}$ and the hybrid kernel \eqref{GPC_Hkernel}. We now perform GP classification using the Laplace approximation.

\subsection{Inference with Laplace approximation}

Given the hybrid kernel $k(\cdot,\cdot)$ defined in \eqref{GPC_Hkernel}, the latent function values follow a Gaussian prior
\begin{equation}
\mathbf{f} \sim \mathcal{N}(\mathbf{0}, K), 
\qquad K_{ij} = k\bigl((\mathbf{x}_i, G_i), (\mathbf{x}_j, G_j)\bigr),
\end{equation}
where $\mathbf{f} = [f(\mathbf{x}_1, G_1),\dots,f(\mathbf{x}_N, G_N)]^\top$ are the latent function values at the training inputs $\mathbf{X} = \{(\mathbf{x}_i, G_i)\}_{i=1}^N$. 
The posterior over $\mathbf{f}$ is analytically intractable:
\begin{equation}
p(\mathbf{f}\mid \mathbf{X}, \mathbf{y}) \propto 
p(\mathbf{y}\mid \mathbf{f}, \mathbf{X}) \, p(\mathbf{f}\mid \mathbf{X}),
\end{equation}
where $p(\mathbf{f}\mid \mathbf{X})$ is the Gaussian prior $\mathcal{N}(\mathbf{f}\mid 0,K)$, so we
employ the Laplace approximation.

The marginal likelihood (or model evidence), conditioned on the hyperparameters $\boldsymbol{\theta}=(\sigma,l,\sigma_g,\ell_g, \lambda)^\top$, is defined by marginalizing over the latent function $\mathbf{f}$:
\begin{equation}
p(\mathbf{y} \mid \mathbf{X}, \boldsymbol{\theta}) = \int p(\mathbf{y}\mid \mathbf{f}, \mathbf{X}) \, p(\mathbf{f} \mid \mathbf{X}, \boldsymbol{\theta}) \, d\mathbf{f}.
\end{equation}

To approximate the posterior $p(\mathbf{f}\mid \mathbf{X}, \mathbf{y})$, we employ the Laplace method, which replaces the posterior with a Gaussian centred at its mode $\mathbf{f}^*$.

The unnormalised log-posterior density is
\begin{equation}
\label{log_p}
\Psi(\mathbf{f}) = \log p(\mathbf{y}\mid \mathbf{f}, \mathbf{X}) - \tfrac{1}{2}\mathbf{f}^\top K^{-1}\mathbf{f}.
\end{equation}
The mode of (\ref{log_p}), denoted $\mathbf{f}^*$, is found via Newton-Raphson iterations, using
\begin{align}
\nabla \Psi(\mathbf{f}) &= \nabla \log p(\mathbf{y}\mid \mathbf{f}, \mathbf{X}) - K^{-1}\mathbf{f}, \\
\nabla\nabla \Psi(\mathbf{f}) &= -W - K^{-1},
\end{align}
where $W = -\nabla\nabla \log p(\mathbf{y}\mid \mathbf{f}, \mathbf{X})$ is diagonal with entries
\begin{equation}
W_{ii} = \sigma(f_i)(1-\sigma(f_i)).
\end{equation}
At the mode $\mathbf{f}^*$, the posterior is approximated as:
\begin{equation}
p(\mathbf{f}\mid \mathbf{X}, \mathbf{y}) \approx 
\mathcal{N}\!\left(\mathbf{f}\mid\mathbf{f}^*, \, (K^{-1}+W)^{-1}\right).
\end{equation}

The Laplace approximation to the log marginal likelihood is
\begin{equation}
\log p(\mathbf{y}\mid \mathbf{X}, \boldsymbol{\theta}) \approx 
-\tfrac{1}{2} \mathbf{f}^{*\top} K^{-1}\mathbf{f}^*
+ \log p(\mathbf{y}\mid \mathbf{f}^*, \mathbf{X})
-\tfrac{1}{2} \log \big| I + W^{1/2} K W^{1/2} \big|,
\label{loglike}
\end{equation}
where $\boldsymbol{\theta}$ denotes the set of all kernel hyperparameters. This quantity is used for hyperparameter optimisation and model comparison.

For Bayesian inference over the kernel hyperparameters $\boldsymbol{\theta}$, we place a prior distribution $p(\boldsymbol{\theta})$ and seek to draw samples from the posterior
\[
p(\boldsymbol{\theta}\mid \mathbf{y},\mathbf{X})
\;\propto\;
p(\mathbf{y}\mid \mathbf{X},\boldsymbol{\theta})\, p(\boldsymbol{\theta}),
\]
where $p(\mathbf{y}\mid \mathbf{X},\boldsymbol{\theta})$ denotes the
Laplace-approximated marginal likelihood obtained from (\ref{loglike}).

As the posterior is not analytically tractable, to facilitate inference, we employ MCMC method that involves sampling via random walk Metropolis (RWM) algorithm \cite{metropolis1953equation, hastings1970monte}. 
For each component $\theta_j$, one swap of RWM operates as follows, in state $\theta_j^{(t)}$, we generate a proposed value $\vartheta_j$ via $\log(\vartheta_j)=\log(\theta_j^{t})+s\epsilon_j$, where $\epsilon_j\sim  \mathcal{N}(0, 1)$  
which is accepted as the next state of the chain, $\theta^{(t+1)}=\vartheta_j$, with probability
\[
\alpha
=
\min\!\left(
1,\;
\frac{
p(\mathbf{y}\mid \mathbf{X},\vartheta)
\; 
}{
p(\mathbf{y}\mid \mathbf{X},\boldsymbol{\theta}^{(t)})
\; 
}
\frac{\prod_j \vartheta_j}{\prod_j \theta_j^{(t)}}
\right),
\]
and rejected, i.e. $\theta^{(t+1)}=\theta^{(t)}$, otherwise. The step size $s$ is adaptively tuned for the algorithm to achieve optimal acceptance probability $0.234$ \cite{gelman1997weak}.

\subsection{Prediction}

Prediction involves computing the posterior predictive probability of a new output $y^*$ given a test input $(\mathbf{x}^*, G^*)$ and the training data $\mathcal{D} = \{\mathbf{X}, \mathbf{G}, \mathbf{y}\}$. This requires marginalising over both the latent function $f^*$ and the kernel hyperparameters $\boldsymbol{\theta}$.

For all datasets we adopt the coding $y\in\{0,1\}$, where $y=1$ denotes the
positive (disease) class and $y=0$ denotes the control class. The full predictive probability of the positive class ($y^*=1$) is defined as:
\begin{equation}
\begin{split}
p(y^*=1 \mid \mathbf{x}^*, G^*, \mathbf{X}, \mathbf{y})
&= \iint p(y^*=1 \mid f^*) \,
p(f^* \mid \mathbf{x}^*, G^*, \mathbf{f}, \mathbf{X}) \\
&\quad \times p(\mathbf{f} \mid \mathbf{X}, \mathbf{y}) \,
df^* \, d\mathbf{f}.
\end{split}
\end{equation}

Under the Laplace approximation, and conditioned on a fixed set of hyperparameters $\boldsymbol{\theta}$, the predictive distribution for the latent value $f^*$ is approximated by a Gaussian:
\begin{equation}
p(f^* \mid \mathbf{x}^*, G^*, \mathbf{X}, \mathbf{y}, \boldsymbol{\theta}) 
\approx \mathcal{N}\!\left(f^* \mid \mu^*, v^*\right),
\end{equation}
with predictive mean $\mu^*$ and variance $v^*$ given by:
\begin{align}
\mu^* &= K_*^\top (K+W^{-1})^{-1} \mathbf{f}^*, \\
v^* &= k\bigl((\mathbf{x}^*,G^*),(\mathbf{x}^*,G^*)\bigr) - K_*^\top (K+W^{-1})^{-1} K_*,
\end{align}
where $K_* = [\,k((\mathbf{x}_1,G_1),(\mathbf{x}^*,G^*)), \dots, k((\mathbf{x}_N,G_N),(\mathbf{x}^*,G^*))\,]^\top$ is the vector of test-train covariances, and $W$ is the Hessian matrix evaluated at the mode $\mathbf{f}^*$.

The predictive probability of the positive class, conditioned on fixed $\boldsymbol{\theta}$, is then obtained by integrating the link function $\sigma(f^*)$ against the approximate Gaussian distribution:
\begin{equation}
p(y^*=1 \mid \mathbf{x}^*, G^*, \mathbf{X}, \mathbf{y}, \boldsymbol{\theta})
\approx \int \sigma(f^*) \,\mathcal{N}(f^* \mid \mu^*, v^*) \, df^*.
\end{equation}

To incorporate uncertainty in the hyperparameter choice, we average the predictive probability across $M$ samples $\{\boldsymbol{\theta}^{(m)}\}_{m=1}^M$ drawn from the hyperparameter posterior $p(\boldsymbol{\theta} \mid \mathbf{X}, \mathbf{y})$ using the RWM algorithm:
\begin{equation}
p(y^*=1 \mid \mathbf{x}^*, G^*, \mathbf{X}, \mathbf{y})
\approx \frac{1}{M}\sum_{m=1}^M 
p(y^*=1 \mid \mathbf{x}^*, G^*, \mathbf{X}, \mathbf{y}, \boldsymbol{\theta}^{(m)}).
\end{equation}

\section{IMBALANCE-HANDLING STRATEGIES}
\label{sec:imbalance}

Microbiome classification problems are often strongly class-imbalanced, with disease-associated samples forming a minority class. In this setting, model training and evaluation can be sensitive to the class proportions: a classifier may achieve high accuracy by favouring the majority class, while performing poorly on the minority class of primary interest. To address this issue within our GP framework, we consider both data-level and decision-level approaches. First, we consider two data-level oversampling approaches (SMOTE and AdaLoRAS) to increase minority representation. Second, we adjust the posterior predictive probabilities to account for  potential shifts in class proportions between the training and test sets.  Using the training confusion matrix and the predicted positive rate on the unlabelled test set, we estimate the test-time class prior and correct the  classifier outputs accordingly.

\subsection{Oversampling}

To address class imbalance inherent in microbiome datasets, where disease-associated samples frequently constitute a minority class, we tested two over-sampling techniques prior to model training. For both methods, all preprocessing (variance thresholding, standardisation) and oversampling are performed strictly within cross-validation training folds to prevent data leakage. 

\begin{itemize}
    \item Synthetic Minority Over-sampling Technique (SMOTE) \cite{chawla2002smote} generates synthetic minority-class samples by interpolating between existing instances in feature space along k-nearest-neighbour edges, thereby augmenting the training distribution without simple duplication and reducing classifier bias toward the majority class.

\item We also developed a hybrid oversampling algorithm optimised for microbiome data by combining principles from ADASYN (Adaptive Synthetic Sampling) \cite{he2008adasyn} and LoRAS (Localized Random Affine Shadowsampling)\cite{bej2021loras}. For each minority-class sample, AdaLoRAS (the hybrid methodology) first computes a difficulty weight proportional to the fraction of its k-nearest neighbours that belong to the majority class. This is motivated by the ADASYN rationale that harder-to-classify boundary samples warrant more synthetic augmentation. The number of synthetic samples to generate per minority instance is allocated proportionally to these difficulty weights. Then synthetic samples are then constructed using a LoRAS-inspired shadow mechanism: for each generation, shadow candidate neighbours are drawn from the minority-only neighbourhood, perturbed with IQR-scaled Gaussian noise to form a cloud of shadow points. Finally, a convex affine combination of randomly selected shadows is computed using Dirichlet-distributed weights. This produces synthetic samples that remain within the local convex hull of the minority neighbourhood while introducing controlled stochastic variation. It also reduces the risk of interpolating across cohort boundaries, since feature values are clipped to the observed training range. Algorithm \ref{alg:adaloras} describes this oversampling strategy.
\end {itemize}

\begin{algorithm}[hbt!]
\caption{AdaLoRAS: Adaptive Synthetic Oversampling for Microbiome Data}
\label{alg:adaloras}
\begin{algorithmic}[1]
\Require
    \Statex $X$: Training feature matrix
    \Statex $y$: Binary class labels (minority class = $1$, majority class = $0$)
    \Statex $k$: Number of nearest neighbours for difficulty estimation
    \Statex $k_{\text{shadow}}$: Number of minority-only neighbours for shadow generation
    \Statex $n_{\text{shadows}}$: Number of shadow candidates per synthetic sample
    \Statex $N$: Total number of synthetic samples to generate

\Statex
\Statex \textbf{Stage 1: Compute Difficulty Weights (ADASYN)}
\State $X_{\text{min}} \gets \{ x_i \in X \mid y_i = 1 \}$
\State $X_{\text{maj}} \gets \{ x_i \in X \mid y_i = 0 \}$
\State $R_{\text{total}} \gets 0$

\For{\textbf{each} $x_i \in X_{\text{min}}$}
    \State $\mathcal{N}_k(x_i) \gets$ find $k$-nearest neighbours of $x_i$ in $X$ 
    \State $r_i \gets \frac{|\{ x_j \in \mathcal{N}_k(x_i) \mid y_j = 0 \}|}{k}$ \Comment{Fraction of majority neighbours}
    \State $R_{\text{total}} \gets R_{\text{total}} + r_i$
\EndFor

\For{\textbf{each} $x_i \in X_{\text{min}}$}
    \State $\Delta_i \gets \frac{r_i}{R_{\text{total}}}$ \Comment{Normalise difficulty weights}
    \State $g_i \gets \lfloor \Delta_i \cdot N \rfloor$ \Comment{Synthetic samples allocated to $x_i$}
\EndFor

\Statex
\Statex \textbf{Phase 2: Generate Synthetic Samples (LoRAS)}
\State $X_{\text{syn}} \gets \emptyset$

\For{\textbf{each} $x_i \in X_{\text{min}}$ \textbf{where} $g_i > 0$}
    \State $\mathcal{N}_{\text{shadow}}(x_i) \gets$ find $k_{\text{shadow}}$-nearest neighbours of $x_i$
    \State Compute $\text{IQR}_f$ for each feature $f$ over $\{ x_i \} \cup \mathcal{N}_{\text{shadow}}(x_i)$
    
    \For{$t = 1$ \textbf{to} $g_i$}
        \Statex \hspace{\algorithmicindent} \textbf{2a: Build shadow cloud}
        
        \For{$j = 1$ \textbf{to} $n_{\text{shadows}}$}
            \State Sample $\varepsilon_{jf} \sim \mathcal{N}(0, \text{IQR}_f^2)$ for each feature $f$
            \State $s_j \gets v_j + \varepsilon_{jf}$ \Comment{IQR-scaled Gaussian perturbation}
        \EndFor
        
        \Statex \hspace{\algorithmicindent} \textbf{2b: Convex affine combination}
        \State Sample $w \sim \text{Dirichlet}(\mathbf{1}_{n_{\text{shadows}}})$,$w_j \ge 0, \sum w_j = 1$
        \State $x_{\text{syn}} \gets \sum_j w_j \cdot s_j$
        
        \Statex \hspace{\algorithmicindent} \textbf{2c: Boundary clipping}
        \For{\textbf{each} feature $f$}
            \State $x_{\text{syn},f} \gets \text{clip}\big(x_{\text{syn},f}, \min_f(X), \max_f(X)\big)$
        \EndFor
        
        \State $X_{\text{syn}} \gets X_{\text{syn}} \cup \{ x_{\text{syn}} \}$
    \EndFor
\EndFor

\State \Return $X_{\text{syn}}$
\end{algorithmic}
\end{algorithm}


\subsection{Prior adjusting using a confusion matrix}
\label{prior}

A Gaussian process classifier produces probabilities for the class label based on the proportion of positives and negatives in the training data. However, the test set may contain a different proportion of positive samples than the training set. When this happens, the probabilities computed by the classifier can become misleading; they will tend to be too large if the test set has fewer positives than the training set, and too small if the test set has more positives. Because the true labels of the test set are not available, we cannot retrain the classifier to match this new situation. Therefore, there is need to adjust the predicted probabilities so that they better reflect the likely proportion of positive samples in the test set. 

We introduce an adjustment method that first summarises how the classifier behaves on the training data using a confusion matrix. We then use the proportion of test points predicted as positive on the unlabelled test set to estimate the test-time positive class proportion. Finally, we adjust the probabilities produced under the training class proportion so that they match the estimated proportion in the test set.

\par
Binary GP classifiers produce posterior predictive probabilities
$p_b(x)=\Pr(y=1\mid x)$, and these probabilities are computed under the
class proportion present in the training set. Let
\[
\pi_b
=\frac{n^{\mathrm{train}}_1}{n^{\mathrm{train}}_0+n^{\mathrm{train}}_1},
\qquad
\pi_t
=\frac{n^{\mathrm{test}}_1}{n^{\mathrm{test}}_0+n^{\mathrm{test}}_1},
\]
where $n^{\mathrm{train}}_j$ and $n^{\mathrm{test}}_j$ count the number of samples with label $y=j$ in the training and test sets, respectively. We refer to $\pi_b$ as the the proportion of samples belonging to the positive class in the training set and to $\pi_t$ as the the proportion of samples belonging to the positive class in the test set. 

\par Given $p_b(x)$, we define the $0/1$ decision at threshold $0.5$:
\begin{equation}
\delta(x)=\ind\{p_b(x)>0.5\},
\end{equation}
where $\ind\{\cdot\}$ is the indicator function. On the training set we know the true label $y\in\{0,1\}$ and we can evaluate $\delta(x)$. We define the $2\times 2$ confusion matrix
\begin{equation}
C(i,j)=\Pr(\delta=i\mid y=j),\qquad i,j\in\{0,1\}.
\end{equation}
This matrix is column-conditional because each column $j$ is normalised to sum to one:
\begin{equation}
C(0,j)+C(1,j)=1.
\end{equation}
In particular, $C(1,1)$ is the true positive rate, $C(0,1)$ is the false negative rate, $C(1,0)$ is the false positive rate, and $C(0,0)$ is the true negative rate. We estimate $C$ by $K$-fold or a held-out split of the training set, counting outcomes for each true class and normalising each column.

On the unlabeled test set the true class labels are unknown, but we can evaluate the hard decisions of the classifier by thresholding each probability $p_b(x)$ at $0.5$ to obtain $\delta(x)=\ind\{p_b(x)>0.5\}$. The empirical frequencies of these decisions define the decision distribution
\[
p(\delta)=
\begin{bmatrix}
\Pr(\delta=0)\\[2pt]
\Pr(\delta=1)
\end{bmatrix}.
\]

\par Assuming the confusion matrix is unchanged between training and test (i.e., the column-normalised entries $C(i,j)=\Pr(\delta=i\mid y=j)$ are the same for both and satisfy $C(0,j)+C(1,j)=1$), the law of total probability gives
\begin{equation}
\label{train-prior}
p(\delta)=
\begin{bmatrix}
\Pr(\delta=0)\\[2pt]
\Pr(\delta=1)
\end{bmatrix}
=
\underbrace{
\begin{bmatrix}
C(0,0) & C{(0,1)}\\
C(1,0) & C{(1,1)}
\end{bmatrix}
}_{C}
\underbrace{
\begin{bmatrix}
\Pr(y=0)\\[2pt]
\Pr(y=1)
\end{bmatrix}
}_{p(y)}.
\end{equation}
We parameterise the test-time class probabilities by the positive-class
proportion $\pi_t$:
\begin{equation}
p(y)=
\begin{bmatrix}
\Pr(y=0)\\[2pt]
\Pr(y=1)
\end{bmatrix}
=
\begin{bmatrix}
1-\pi_t\\[2pt]
\pi_t
\end{bmatrix},
\end{equation}
which is unknown.

In \eqref{train-prior}, $\Pr(\delta=1)$) can be written explicitly as
\begin{equation}
\begin{aligned}
\Pr(\delta=1)
&= C(1,0)\Pr(y=0) + C(1,1)\Pr(y=1) \\
&= C(1,0)\bigl(1-\pi_t\bigr) + C(1,1)\pi_t .
\end{aligned}
\end{equation}
Expanding and grouping terms in $\pi_t$ gives
\begin{align}
\Pr(\delta=1)
&= C(1,0) - C(1,0)\pi_t + C(1,1)\pi_t \\
&= C(1,0) + \bigl(C(1,1)-C(1,0)\bigr)\,\pi_t.
\end{align}
Solving this expression for $\pi_t$ yields
\begin{equation}
\pi_t
=
\frac{\Pr(\delta=1)-C(1,0)}{C(1,1)-C(1,0)}.
\end{equation}
In practice, we replace probabilities by their empirical estimates from the
test decisions, leading to
\begin{equation}
\widehat{\pi}_t
=
\frac{\Pr(\delta=1)-C(1,0)}{C(1,1)-C(1,0)}.
\end{equation}
This is the closed-form solution of the $2\times 2$ linear system
$p(\delta)=C\,p(y)$ in the binary case.

The procedure for estimating the test class proportion from the observed decision rate and the confusion matrix is a standard technique in prior shift adaptation \cite{saerens2002adjusting}.

\par
To adjust the posterior probabilities to a different class prior, we apply a standard result from Bayes' rule. For any prior
$\pi=\Pr(y=1)$, the posterior probability of the positive class is
\[
\Pr(y=1\mid x)
=
\frac{p(x\mid y=1)\,\Pr(y=1)}
     {p(x\mid y=1)\Pr(y=1) + p(x\mid y=0)\Pr(y=0)},
\]
and similarly
\[
\Pr(y=0\mid x)
=
\frac{p(x\mid y=0)\,\Pr(y=0)}
     {p(x\mid y=1)\Pr(y=1) + p(x\mid y=0)\Pr(y=0)}.
\]
Taking the ratio of these two expressions cancels the common denominator,
which gives the posterior odds in terms of the prior odds and the likelihood
ratio:
\begin{equation}
\frac{\Pr(y=1\mid x)}{\Pr(y=0\mid x)}
=
\frac{\Pr(y=1)}{\Pr(y=0)}
\cdot
\frac{p(x\mid y=1)}{p(x\mid y=0)}.
\label{eq:posterior_odds_general}
\end{equation}
Since $\Pr(y=1)=\pi$ and $\Pr(y=0)=1-\pi$, the prior odds are
$\pi/(1-\pi)$, and \eqref{eq:posterior_odds_general} becomes
\begin{equation}
\frac{\Pr(y=1\mid x)}{\Pr(y=0\mid x)}
=
\frac{\pi}{1-\pi}\,
\frac{p(x\mid y=1)}{p(x\mid y=0)}.
\label{eq:posterior_odds_specific}
\end{equation}

\par
Let the prior used during training be $\pi_b$, from \eqref{eq:posterior_odds_specific}, the posterior odds under the training
prior are
\begin{equation}
O_b(x)
=
\frac{p_b(x)}{1-p_b(x)}
=
\frac{\pi_b}{1-\pi_b}\,
\frac{p(x\mid y=1)}{p(x\mid y=0)}.
\label{eq:Ob}
\end{equation}
If we change only the prior to $\pi_t$ (the test-time prior), the posterior
odds become
\begin{equation}
O_t(x)
=
\frac{\pi_t}{1-\pi_t}\,
\frac{p(x\mid y=1)}{p(x\mid y=0)}.
\label{eq:Ot}
\end{equation}
Dividing \eqref{eq:Ot} by \eqref{eq:Ob}
gives a simple multiplicative relation:
\begin{equation}
\frac{O_t(x)}{O_b(x)}
=
\frac{\pi_t(1-\pi_b)}{\pi_b(1-\pi_t)}
\;\equiv\;
\kappa,
\end{equation}
so that
\begin{equation}
O_t(x)=\kappa\,O_b(x),
\qquad
\kappa=\frac{\pi_t(1-\pi_b)}{\pi_b(1-\pi_t)}.
\end{equation}
To obtain the corrected probability under the test prior, we convert odds
back into probability:
\begin{equation}
p_t(x)
=
\frac{O_t(x)}{1+O_t(x)}
=
\frac{\kappa\,p_b(x)}{1-p_b(x)+\kappa\,p_b(x)}.
\end{equation}
In practice, the unknown test prior $\pi_t$ is replaced by its estimate
$\widehat{\pi}_t$ obtained in the previous step.

However, in applications, the cost of misclassification is often asymmetric. Specifically, the false positive cost $c_{10}$ (cost incurred when $y=0$ is predicted as $\hat{y}=1$) and the false negative cost $c_{01}$ (cost incurred when $y=1$ is predicted as $\hat{y}=0$) may differ significantly. To minimise the overall expected loss, Bayesian decision theory dictates that the optimal threshold $\tau^\star$ is determined by the ratio of these misclassification costs. The decision rule is to predict $\hat{y}=1$ if the corrected posterior probability $p_t(\mathbf{x})$ exceeds $\tau^\star$.

The optimal threshold $\tau^\star$ is derived by minimising the expected loss, which is equivalent to predicting $\hat{y}=1$ when the posterior odds $O_t(\mathbf{x})$ exceeds the cost ratio $c_{01}/c_{10}$. This leads to:
\begin{equation}
\tau^\star=\frac{c_{01}}{c_{01}+c_{10}},
\label{eq:cost_threshold}
\end{equation}
where the final prediction is
\begin{equation}
\hat y(\mathbf{x})=\ind\{p_t(\mathbf{x})>\tau^\star\}.
\end{equation}
This allows the final classifier to be optimally tuned to the specific costs of the application \cite{elkan2001foundations}. For instance, if the false negative cost $c_{01}$ is significantly higher than $c_{10}$ , the optimal threshold
$\tau^\star$ will be low. This makes the classifier more sensitive, that is,
more likely to predict $\hat{y}=1$.


\section{NUMERICAL RESULTS AND DISCUSSION}
\label{sec:results}
\subsection{Datasets}

Data for this analysis were obtained from the Borenstein lab curated gut microbiome-metabolome dataset collection, a publicly available centralized resource of paired human fecal microbiome and metabolome profiles \cite{muller2022gut}. This repository provides uniformly processed datasets across multiple independent single-cell studies; it provides genus-level taxonomic abundance tables, metabolite abundance tables, metabolite identifier mapping tables, and associated sample metadata.

For this work, we selected three microbiome datasets from this collection to test the methodology on various immunological responses and dataset cohort balances. The selected datasets are summarized in Table \ref{tab:selected_cohorts}. The datasets were preprocessed using standard bioinformatic pipelines \cite{muller2022gut}. When multiple diagnostic categories were available, the problem was recast as a binary classification task. 

\begin{table}[htpb]
\centering
\caption{Summary of Selected Cohorts}
\label{tab:selected_cohorts}
\begin{tabular}{p{0.2\linewidth} p{0.3\linewidth} c c}
\hline
\textbf{Name} & \textbf{Description} & \textbf{Longitudinal} & \textbf{Samples} \\
\hline
KIM \cite{kim2020fecal} & Patients with advanced colorectal adenomas, CRC, and controls & No & 240 \\
KOSTIC \cite{kostic2015dynamics} & Children at risk for T1D & Yes & 103  \\
SINHA \cite{sinha2016fecal} & CRC patients and controls & No & 131 \\
\hline
\end{tabular}
\end{table}

\subsection{Baseline methods}

To benchmark our approach, we compare our proposed GP method against a set of standard classifiers commonly used in high-dimensional microbiome classification. The baseline methods operate only on the vector features and do not use any graph information.

We consider five representative classifiers: Gaussian Naive Bayes, linear support vector machines \cite{cortes1995support}, logistic regression, random forests \cite{breiman2001random}, and $k$-nearest neighbours \cite{cover1967nearest}. Each model is trained on the same vector representation used for the vector component of our hybrid kernel, with the same preprocessing and normalisation applied across methods.

To account for the presence of class imbalance in several datasets, we evaluate these baselines under three strategies: no resampling, SMOTE, and AdaLoRAS. Tables~\ref{tab:baseline_kim}--\ref{tab:baseline_sinha} summarise baseline performance on each dataset. To keep the comparison concise, for each classifier we report the best mean F1-score over the three imbalance settings, together with the corresponding accuracy, precision, and recall. 

\begin{table}[t]
\centering
\caption{Baseline performance on the \textit{kim} dataset. For each classifier we report the best mean F1-score over \{no resampling, SMOTE, AdaLoRAS\} (mean $\pm$ std across folds).}
\label{tab:baseline_kim}
\footnotesize
\setlength{\tabcolsep}{4pt}
\renewcommand{\arraystretch}{0.98}
\begin{tabular}{p{2.2cm}cccc}
\toprule
\textbf{Baseline} & \textbf{Accuracy} & \textbf{Precision} & \textbf{Recall} & \textbf{F1-score} \\
\midrule
Gaussian NB (SMOTE) & 0.492 $\pm$ 0.049 & 0.562 $\pm$ 0.037 & 0.515 $\pm$ 0.093 & 0.535 $\pm$ 0.062 \\
Linear SVM & 0.571 $\pm$ 0.017 & 0.578 $\pm$ 0.011 & 0.934 $\pm$ 0.044 & 0.714 $\pm$ 0.021 \\
Logistic regression & 0.562 $\pm$ 0.048 & 0.617 $\pm$ 0.043 & 0.638 $\pm$ 0.048 & 0.626 $\pm$ 0.037 \\
Random forest & 0.571 $\pm$ 0.034 & 0.589 $\pm$ 0.022 & 0.841 $\pm$ 0.070 & 0.692 $\pm$ 0.032 \\
$k$NN & 0.550 $\pm$ 0.045 & 0.576 $\pm$ 0.034 & 0.833 $\pm$ 0.068 & 0.680 $\pm$ 0.036 \\
\bottomrule
\end{tabular}
\end{table}

\begin{table}[t]
\centering
\caption{Baseline performance on the \textit{kostic} dataset. For each classifier we report the best mean F1-score over \{no resampling, SMOTE, AdaLoRAS\} (mean $\pm$ std across folds).}
\label{tab:baseline_kostic}
\footnotesize
\setlength{\tabcolsep}{4pt}
\renewcommand{\arraystretch}{0.98}
\begin{tabular}{p{2.5cm}cccc}
\toprule
\textbf{Baseline} & \textbf{Accuracy} & \textbf{Precision} & \textbf{Recall} & \textbf{F1-score} \\
\midrule
Gaussian NB & 0.735 $\pm$ 0.120 & 0.815 $\pm$ 0.102 & 0.769 $\pm$ 0.175 & 0.781 $\pm$ 0.111 \\
Linear SVM & 0.728 $\pm$ 0.059 & 0.711 $\pm$ 0.039 & 0.969 $\pm$ 0.038 & 0.820 $\pm$ 0.039 \\
Logistic regression & 0.796 $\pm$ 0.081 & 0.836 $\pm$ 0.083 & 0.862 $\pm$ 0.132 & 0.840 $\pm$ 0.076 \\
Random forest (AdaLoRAS) & 0.788 $\pm$ 0.069 & 0.778 $\pm$ 0.050 & 0.942 $\pm$ 0.084 & 0.850 $\pm$ 0.050 \\
$k$NN & 0.690 $\pm$ 0.048 & 0.683 $\pm$ 0.037 & 0.969 $\pm$ 0.038 & 0.800 $\pm$ 0.028 \\
\bottomrule
\end{tabular}
\end{table}

\begin{table}[t]
\centering
\caption{Baseline performance on the \textit{sinha} dataset. For each classifier we report the best mean F1-score over \{no resampling, SMOTE, AdaLoRAS\} (mean $\pm$ std across folds).}
\label{tab:baseline_sinha}
\footnotesize
\setlength{\tabcolsep}{4pt}
\renewcommand{\arraystretch}{0.98}
\begin{tabular}{p{2.2cm}cccc}
\toprule
\textbf{Baseline} & \textbf{Accuracy} & \textbf{Precision} & \textbf{Recall} & \textbf{F1-score} \\
\midrule
Gaussian NB & 0.503 $\pm$ 0.084 & 0.314 $\pm$ 0.060 & 0.456 $\pm$ 0.143 & 0.363 $\pm$ 0.082 \\
Linear SVM (SMOTE) & 0.572 $\pm$ 0.078 & 0.330 $\pm$ 0.143 & 0.308 $\pm$ 0.117 & 0.316 $\pm$ 0.124 \\
Logistic regression (AdaLoRAS) & 0.572 $\pm$ 0.018 & 0.344 $\pm$ 0.032 & 0.386 $\pm$ 0.106 & 0.360 $\pm$ 0.063 \\
Random forest (SMOTE) & 0.626 $\pm$ 0.042 & 0.321 $\pm$ 0.181 & 0.214 $\pm$ 0.140 & 0.252 $\pm$ 0.153 \\
$k$NN (AdaLoRAS) & 0.411 $\pm$ 0.064 & 0.318 $\pm$ 0.039 & 0.717 $\pm$ 0.089 & 0.438 $\pm$ 0.042 \\
\bottomrule
\end{tabular}
\end{table}

\subsection{Results}
We evaluated our proposed hybrid Gaussian process kernel on several microbiome datasets, where each sample is represented by both vector features (e.g., abundance profiles) and a sample-specific graph. 

Tables~\ref{tab:kostic_results}--\ref{tab:sinha_results} summarise cross-validated performance (mean $\pm$ standard deviation). Because the cohorts are imbalanced, we interpret accuracy with caution and emphasise minority-class recall and F1-score, which better reflect the ability to identify positive cases.

We observed that incorporating graph structure can improve performance, but the effect depends on both the graph construction and the imbalance-handling strategy. On \textit{sinha}, the largest improvement in minority-class performance is obtained when graph information is combined with oversampling: the graph-based GP with SMOTE achieves the highest mean F1-score (0.733), whereas the vector-only GP without oversampling attains substantially lower F1 values (Table~\ref{tab:sinha_results}). On \textit{kim}, the best F1-score is also obtained with a graph-based model paired with SMOTE and adaptive thresholding (F1 = 0.717; Table~\ref{tab:kim_results}), suggesting that structural similarity can be beneficial when the decision rule is tuned for imbalance.

For the \textit{kim} dataset,the best-performing graph-based configuration is the bootstrap-filtered kernel with SMOTE and adaptive thresholding (Table~\ref{tab:kim_results}), achieving F1 = 0.717 with recall = 0.961 and precision = 0.572. This is comparable to the strongest vector-only baseline (Linear SVM; Table~\ref{tab:baseline_kim}), while providing a probabilistic framework that supports uncertainty-aware analysis.

The benefit of incorporating topological prior knowledge is most pronounced on the \textit{sinha} cohort. Here, the bootstrap-filtered graph kernel combined with SMOTE attains the highest mean F1-score of 0.733 (precision = 0.858, recall = 0.642; Table~\ref{tab:sinha_results}). This substantially outperforms the best-performing vector-only baseline ($k$NN; F1 = 0.438, Table~\ref{tab:baseline_sinha}), demonstrating that structural information provides a critical signal when vector features alone are insufficient.

Finally, on \textit{kostic}, several unstructured baselines inherently perform well (Table~\ref{tab:baseline_kostic}). Within our GP framework, oversampling dramatically improves minority-class detection, and the graph-augmented model remains highly competitive (e.g., bootstrap-filtered with SMOTE yields F1 = 0.801, recall = 0.962; Table~\ref{tab:kostic_results}). Nonetheless, the empirical results suggest that the relative performance gains from graph embeddings are highly dataset-dependent, with the most significant advantages materialising in cohorts like \textit{sinha} and \textit{kim}.

\begin{table*}[ht]
\centering
\caption{Performance comparison on the \textit{kim} dataset using cross-validation. Results are reported as mean $\pm$ standard deviation over available folds.}
\label{tab:kim_results}
\footnotesize
\setlength{\tabcolsep}{4pt}
\renewcommand{\arraystretch}{0.95}
\begin{tabular}{p{6.2cm}cccc}
\toprule
\textbf{Model} & \textbf{Accuracy} & \textbf{Precision} & \textbf{Recall} & \textbf{F1-score} \\
\midrule
GP (no graphs)
& 0.495 $\pm$ 0.064
& 0.494 $\pm$ 0.056
& 0.640 $\pm$ 0.330
& 0.562 $\pm$ 0.092 \\

GP (no graphs, adaptive threshold)
& 0.633 $\pm$ 0.049
& 0.615 $\pm$ 0.047
& 0.807 $\pm$ 0.092
& 0.693 $\pm$ 0.021 \\

\midrule
GP + graphs (bootstrap filtered)
& 0.564 $\pm$ 0.064
& 0.583 $\pm$ 0.106
& 0.598 $\pm$ 0.112
& 0.585 $\pm$ 0.080 \\

GP + graphs (bootstrap filtered, adaptive threshold)
& 0.574 $\pm$ 0.028
& 0.575 $\pm$ 0.021
& 0.635 $\pm$ 0.359
& 0.556 $\pm$ 0.191 \\

GP + graphs (bootstrap filtered + SMOTE)
& 0.594 $\pm$ 0.036
& 0.615 $\pm$ 0.066
& 0.654 $\pm$ 0.170
& 0.618 $\pm$ 0.076 \\

GP + graphs (bootstrap filtered + SMOTE, adaptive threshold)
& 0.614 $\pm$ 0.048
& 0.572 $\pm$ 0.042
& 0.961 $\pm$ 0.036
& 0.717 $\pm$ 0.038 \\

GP + graphs (LIONESS + SMOTE, adaptive threshold)
& 0.520 $\pm$ 0.074
& 0.498 $\pm$ 0.058
& 0.826 $\pm$ 0.159
& 0.613 $\pm$ 0.057 \\
\bottomrule
\end{tabular}
\end{table*}

\begin{table*}[ht]
\centering
\caption{Performance comparison on the \textit{kostic} dataset using 10-fold cross-validation. Results are reported as mean $\pm$ standard deviation.}
\label{tab:kostic_results}
\footnotesize
\setlength{\tabcolsep}{4pt}
\renewcommand{\arraystretch}{0.95}
\begin{tabular}{p{6.2cm}cccc}
\toprule
\textbf{Model} & \textbf{Accuracy} & \textbf{Precision} & \textbf{Recall} & \textbf{F1-score} \\
\midrule
GP (no graphs)
& 0.613 $\pm$ 0.112
& 0.275 $\pm$ 0.071
& 0.444 $\pm$ 0.160
& 0.325 $\pm$ 0.067 \\

GP (no graphs + SMOTE)
& 0.800 $\pm$ 0.145
& 0.778 $\pm$ 0.166
& 0.911 $\pm$ 0.080
& 0.825 $\pm$ 0.102 \\

\midrule
GP + graphs (bootstrap filtered)
& 0.781 $\pm$ 0.089
& 0.393 $\pm$ 0.245
& 0.420 $\pm$ 0.377
& 0.492 $\pm$ 0.227 \\

GP + graphs (confusion-matrix bootstrap filtered)
& 0.784 $\pm$ 0.053
& 0.791 $\pm$ 0.300
& 0.171 $\pm$ 0.242
& 0.438 $\pm$ 0.128 \\

GP + graphs (bootstrap filtered + SMOTE)
& 0.795 $\pm$ 0.063
& 0.695 $\pm$ 0.108
& 0.962 $\pm$ 0.058
& 0.801 $\pm$ 0.066 \\

\bottomrule
\end{tabular}
\end{table*}

\begin{table*}[t]
\centering
\caption{Performance comparison on the \textit{sinha} dataset using cross-validation. Results are reported as mean $\pm$ standard deviation.}
\label{tab:sinha_results}
\footnotesize
\setlength{\tabcolsep}{4pt}
\renewcommand{\arraystretch}{0.95}
\begin{tabular}{p{6.2cm}cccc}
\toprule
\textbf{Model} & \textbf{Accuracy} & \textbf{Precision} & \textbf{Recall} & \textbf{F1-score} \\
\midrule
GP (no graphs)
& 0.767 $\pm$ 0.049
& 0.260 $\pm$ 0.094
& 0.527 $\pm$ 0.153
& 0.332 $\pm$ 0.087 \\

GP (no graphs + SMOTE, adaptive threshold)
& 0.536 $\pm$ 0.043
& 0.536 $\pm$ 0.043
& 1.000 $\pm$ 0.000
& 0.697 $\pm$ 0.037 \\

\midrule
GP + graphs (bootstrap filtered)
& 0.836 $\pm$ 0.059
& 0.323 $\pm$ 0.114
& 0.265 $\pm$ 0.103
& 0.282 $\pm$ 0.085 \\

GP + graphs (confusion-matrix bootstrap filtered)
& 0.887 $\pm$ 0.047
& 0.513 $\pm$ 0.338
& 0.157 $\pm$ 0.158
& 0.320 $\pm$ 0.097 \\

GP + graphs (bootstrap filtered + SMOTE)
& 0.760 $\pm$ 0.068
& 0.858 $\pm$ 0.049
& 0.642 $\pm$ 0.085
& 0.733 $\pm$ 0.068 \\

\bottomrule
\end{tabular}
\end{table*}

We present calibration and uncertainty analyses on the \textit{sinha} cohort as a representative example, since it exhibits the clearest improvement from incorporating graph structure under class imbalance.

Table~\ref{tab:sinha_calibration} reports three complementary measures of probabilistic quality for the two graph constructions, averaged across held-out folds. Lower ECE indicates that predicted probabilities better match empirical outcome frequencies \cite{guo2017calibration}; lower Brier score and NLL indicate more accurate and better-scored probabilistic predictions \cite{niculescu2005predicting}. The \texttt{bootstrap\_filtered} construction yields consistently better calibration across all three metrics (ECE: 0.129 vs 0.138; Brier: 0.160 vs 0.185; NLL: 0.478 vs 0.547), suggesting that bootstrap-filtered graphs provide a more reliable similarity structure for the GP kernel.

\begin{table}[t]
\centering
\caption{Calibration metrics on the \textit{sinha} dataset for the graph-based GP with oversampling, comparing two graph constructions. Values are reported as mean $\pm$ standard deviation across splits ($n=10$). Lower is better for all metrics.}
\label{tab:sinha_calibration}
\setlength{\tabcolsep}{5pt}
\renewcommand{\arraystretch}{1.05}
\begin{tabular}{lccc}
\toprule
\textbf{Graph construction} & \textbf{ECE}  & \textbf{Brier}  & \textbf{NLL}  \\
\midrule
bootstrap\_filtered & 0.129 $\pm$ 0.039 & 0.160 $\pm$ 0.020 & 0.478 $\pm$ 0.042 \\
pearson\_lioness    & 0.138 $\pm$ 0.039 & 0.185 $\pm$ 0.013 & 0.547 $\pm$ 0.028 \\
\bottomrule
\end{tabular}
\end{table}

\begin{figure}[t]
  \centering
  \includegraphics[width=\linewidth]{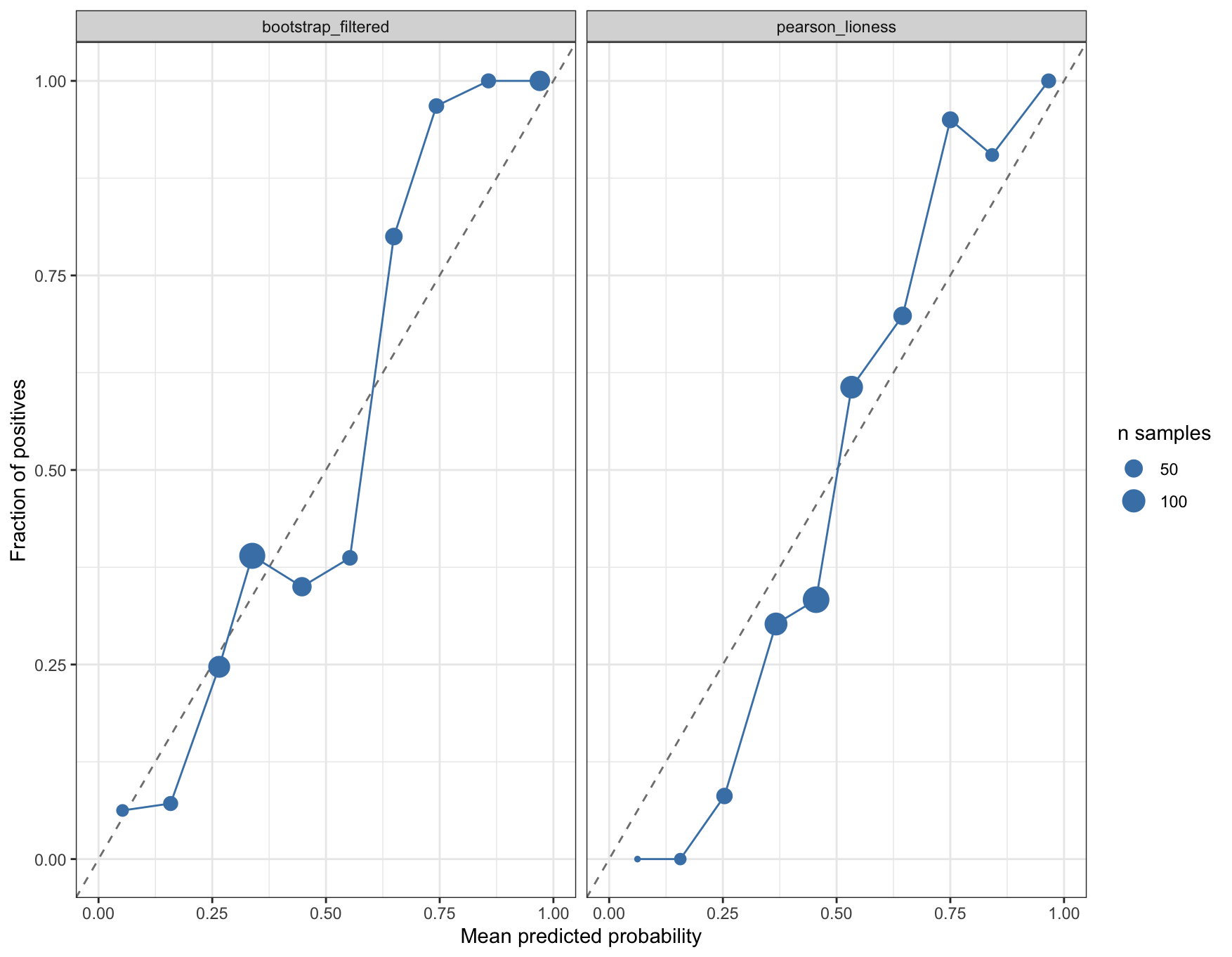}
  \caption{Reliability diagrams on the \textit{sinha} cohort for the graph-based GP
  with oversampling under two graph constructions (\texttt{bootstrap\_filtered} and
  \texttt{pearson\_lioness}). The dashed line indicates perfect calibration; marker
  size reflects the number of samples per bin.}
  \label{fig:reliability}
\end{figure}

In Figure~\ref{fig:reliability}, both constructions deviate from the diagonal, indicating imperfect calibration. For \texttt{bootstrap\_filtered}, predicted probabilities are slightly conservative at the low end, while higher probabilities (above $\approx 0.6$) tend to overestimate the empirical positive rate. The \texttt{pearson\_lioness} variant is reasonably calibrated in the mid-range but shows larger variability at high predicted probabilities, consistent with its higher ECE and NLL.

Across the three datasets, the graph-augmented GP classification yields the most pronounced gains on \textit{sinha}, confirming that structured biological similarity provides critical complementary information beyond raw abundance features. On \textit{kim} and \textit{kostic}, the framework remains highly competitive with strong vector-only baselines, indicating that the magnitude of benefit from topological embeddings is inherently dataset-dependent. Calibration results on \textit{sinha} further show that the predicted probabilities are reasonably well-calibrated, supporting their use for uncertainty-aware decision making.

\subsection{Discussion}
The dataset-dependent performance of the proposed methodology is consistent with known biological differences in underlying signals reported in the original studies. In the Kostic \cite{kostic2015dynamics} study, colorectal tumors show strong enrichment of the bacterium \textit{Fusobacterium nucleatum}. As a result, classification can be achieved using simple abundance differences captured by vector based models, as shown in our experiments. 

In contrast, the Sinha study \cite{sinha2016fecal} highlights extensive correlations between microbial taxa and metabolites associated with colorectal cancer risk, indicating that disease might be linked to coordinated changes in community function. In such settings, modeling microbial interactions, as encoded in our graph kernels, becomes important to reflect disease-associated dysbiosis. 

The Kim dataset \cite{kim2020fecal} appears to fall between these two extremes, with the study reporting differences in composition across cohorts, but the signal is spread across many moderate microbiome-metabolome changes. This likely reflects the heterogeneity of dysbiosis, and we see in our results that helping the minority class occupy feature space more densely improves boundary estimation more strongly. 

Together, these observations suggest that the effectiveness of graph-based representations depends strongly on the biological structure of the microbiome signal, with the largest gains arising in cohorts where disease is characterised by community-level dysbiosis rather than isolated taxonomic markers.

\section{Conclusion}
\label{sec:conclusion}
We proposed a structured Gaussian process classification
framework for high-dimensional, small-sample microbiome data. By incorporating graph-encoded biological information into the kernel, the model combines abundance features with graph-based similarity.

Across three cohorts, the graph-augmented GP achieves competitive performance, with the clearest improvement in minority-class detection on the \textit{sinha} dataset. In addition, calibration results on \textit{sinha} suggest that the predicted probabilities are reasonably well-calibrated, supporting their use in probabilistic decision making.

A limitation of the current framework is the computational cost of GP inference and hyperparameter learning, especially in high-dimensional settings, and the fact that results can be sensitive to kernel choices and imbalance-handling settings. In future work, we will study more scalable GP inference and improved calibration and explore deep GP extensions to learn richer kernel representations from graph-structured data and other omics datasets.

\bibliographystyle{unsrt}  
\bibliography{Reference}  

\end{document}